\documentclass[9pt,twocolumn,twoside]{osajnl}

\usepackage{amsmath,amssymb,graphicx}
\usepackage{verbatim}   
\usepackage[version=3]{mhchem} 

\journal{josab} 

\setboolean{shortarticle}{false}

\title{Frequency stabilization of a 739 nm laser to an $I_2$ spectrum for trapped Ytterbium ions}

\author[1,$\dagger$]{Hao Wu}
\author[1,$\dagger$]{Pengfei Lu}
\author[1,2,$\dagger$]{Yang Liu}
\author[1]{Jiangyong Hu}
\author[1]{Qifeng Lao}
\author[1]{Xinxin Rao}

\author[5]{Lunhua Deng}
\author[1,2,3,*]{Feng Zhu}
\author[1,2,3,4,*]{Le Luo}

\affil[1]{School of Physics and Astronomy,  Sun Yet-sen University, Zhuhai 519082, China}
\affil[2]{Center of Quantum Information Technology, Shenzhen Research Institute of Sun Yat-sen University, Nanshan Shenzhen 518087, China}
\affil[3]{State Key Laboratory of Optoelectronic Materials and Technologies, Sun Yat-sen University( Guang zhou Campus), Guangzhou 510275, China}
\affil[4]{International Quantum Academy (SIQA) and Shenzhen Branch, Hefei National Laboratory, Futian District, Shenzhen, P.R. China}
\affil[5]{State Key Laboratory of Precision Spectroscopy, East China Normal University, Shanghai, 200241, China}
\affil[*]{zhufeng25@mail.sysu.edu.cn}
\affil[*]{luole5@mail.sysu.edu.cn}
\affil[$\dagger$]{These authors contribute equally to this work.}

\begin{abstract}
We report on the frequency stabilization of a 739 nm Ti:sapphire laser to a hyperfine component of the $^{127}I_{2}$ B(1)-X(11) P(70) transition using acousto-optic modulation transfer spectroscopy (MTS). A frequency stability of $3.83\times 10^{-11}$ around 13 s averaging time is achieved when the laser frequency is stabilized. The observed hyperfine transition of the molecular iodine is an ideal frequency reference for locking the lasers used in experiments with trapped ytterbium ions, since its second harmonic frequency is the $^{2}S_{\frac{1}{2}}-^{2}P_{\frac{1}{2}}$ transition of the ytterbium ion at 369.5 nm. By investigating the line broadening effects due to the iodine vapor pressure and laser power, the locking is optimized to the theoretical signal to noise ratio (TSNR) of this iodine transition. 
\end{abstract}

\setboolean{displaycopyright}{true}

\begin{document}

\maketitle

\section{Introduction}
Stable laser sources with narrow linewidth and long-term stability are of great importance for experiments in many fields of physics. These are realized by active frequency stabilization to atomic or molecular references with high resolution Sub-Doppler spectroscopy. In particular, molecular iodine has more than 20,000 hyperfine transitions covering wavelengths from 500 to 780 nm, many of which have been thoroughly investigated, allowing their frequencies to be accurately determined with uncertainties at the kHz level. These frequencies are often used as frequency references for metrology applications or laser frequency stabilization in the visible and near-infrared regions due to their excellent properties. Stabilization of lasers to various hyperfine components of molecular iodine have been demonstrated for many different wavelengths, such as 532 nm\cite{hall1999stabilization}, 548 nm\cite{hsiao2013absolute}, 561 nm\cite{yang2012hyperfine}, 578 nm\cite{kobayashi2016absolute}, 605 nm\cite{bertinetto1987helium}, 612 nm\cite{cerez1979helium}, 633 nm\cite{lazar2000tunable}, 650 nm\cite{xie2019frequency}, 660 nm\cite{guo2004frequency}, 739 nm\cite{olmschenk2007manipulation}, 830 nm\cite{ludvigsen1992frequency}. Some of its hyperfine components are recommended as the frequency standards for the practical realization of the metre \cite{felder2005practical}. Doppler-free spectroscopy, including saturated absorption spectroscopy \cite{preston1996doppler} and modulation transfer spectroscopy \cite{shirley1982modulation}, provides robust frequency stabilization due to the elimination of frequency and amplitude noises from the Doppler background. It has been widely used in laser stabilization systems, where the width of the error signal can be compressed to tens of MHz.

The $Yb^{+}$ ion has been the primary candidate for realization of optical clock\cite{huntemann2016single}, quantum computation \cite{haffner2008quantum} and quantum information processing\cite{wineland2003quantum, wineland2011quantum}. Its ground state (6s) $^{2}S_{1/2}$ and first excited electronic state (6p) $^{2}P_{1/2}$ form a nearly closed cooling cycle with a transition wavelength of 369.5 nm, which corresponds to the second harmonics of iodine absorption lines at 739 nm. Therefore, these iodine lines can be frequency references for research on trapped ytterbium ions \cite{olmschenk2007manipulation}. Previously, a 739 nm diode laser was locked to a reference cavity using Pound-Drever-Hall scheme, while this cavity was locked to the iodine cell using saturated-absorption spectroscopy and modulation transfer technique for keeping its long-term frequency stability \cite{olmschenk2007manipulation}. Here, we further study the iodine spectrum at 739 nm and found the transition used in reference \cite{olmschenk2007manipulation} is B(1)-X(11) R(78). By analyzing the TSNR limit of the slightly stronger adjacent transition of B(1)-X(11) P(70), we found the previous locking scheme can be significantly simplified when a Ti:sapphire laser is used to generate the 739 nm wavelength for trapped ion experiments. Indebted to the short-term stability of the Ti:Sapphire laser, it can be directly locked to the iodine transition to meet the requirement of the trapped ion quantum information experiments. By investigating the line broadening effects due to iodine vapor pressure and laser power, the TSNR of the locking at this transition is optimized and the laser is stabilized at tens of kHz in minutes for the cooling and detecting the Yb ion qubit.

\section{Experiment}

\begin{figure}[H]
\begin{center}
\includegraphics[width=1.0\linewidth]{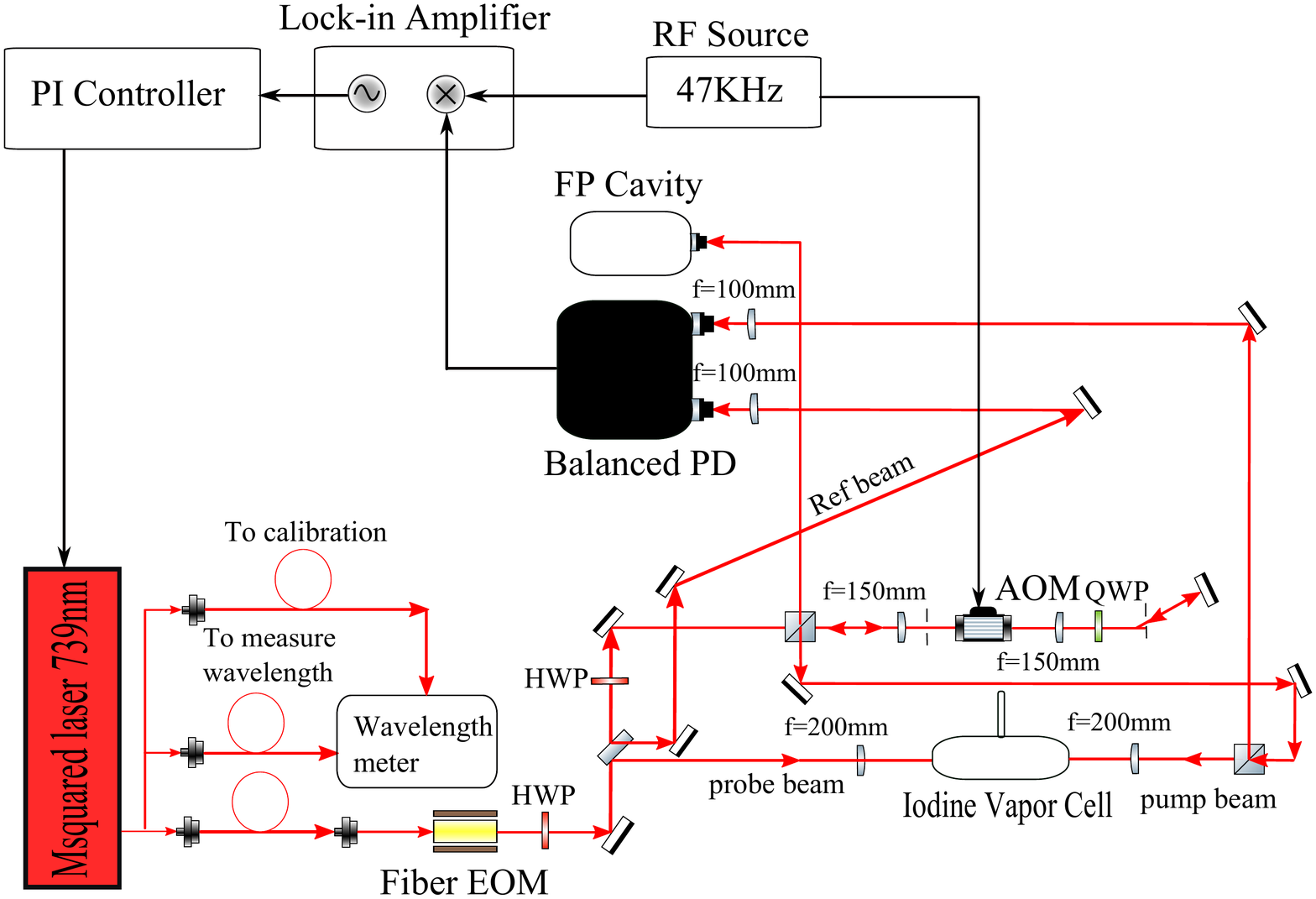}
\caption{Experimental layout for 739 nm laser frequency stabilization using modulation transfer spectroscopy of molecular iodine. EOM: electric-optic modulator, PD: photodiode, FP cavity: Fabry-Parot cavity, HWP: half wave plate, QWP: quarter wave plate. }
\label{setup}
\end{center}
\end{figure}

A schematic of our experimental setup is shown in Figure ~\ref{setup}. The 739 nm light is generated by a Ti: Sapphire laser (SolsTis 4000 PSF XF, M Squared) which is pumped by a diode-pumped solid-state laser (Sprout G15, Lighthouse). It is then frequency doubled by a lithium triborate (\ce{LiB3O5}) crystal in an ECD-F second-harmonic generation box (M Squared) to generate 369.5 nm laser for photo-ionization of ytterbium atoms and Doppler cooling as well as quantum state manipulation of ytterbium ion using the $^{2}S_{1/2}-^{2}P_{1/2}$ transition. About 10 $\mu W$ output 739 nm light is sent to a wavelength meter (WS 7-30, HighFinesse) for real-time wavelength monitoring. Another 25 mW 739.05 nm light is coupled to a fiber electro-optic modulator (fiber-EOM, NIR-MPX800, Ixblue) through a polarization-maintaining optical fiber. Only about 8.0 mW of this light is transmitted because of the coupling and transmission losses. This fiber-EOM has a first-order diffraction efficiency about 30$\% $ as measured by a Fabry-Parot cavity, thus effectively only about 2.4 mW of the first-order sideband is used for the laser stabilization. The transmitted light is then divided into the pump beam, the probe beam and the reference beam by a 92:4:4 beam splitter.

The pump beam passes through an acousto-optic modulator (TEM 200-50-739, Brimrose) using a double-pass scheme, in order to remove the directional jitter of the first-order diffraction light due to the frequency modulation and additional amplitude modulation caused by the power change of the pump beam. The modulation frequency is 47 kHz and the modulation depth is 2 MHz. The power of the probe beam including the carrier and the sidebands is roughly 320 $\mu$W, and the power ratio of the probe beam to the pump beam is adjusted to approximately 1:9. Both the pump beam and the probe beam are focused to about 100$ \mu$m beam waists using $f=200$ mm lenses, and two focal points are tuned to overlap with each other. After passing through the iodine cell, the probe beam is aligned to enter a balanced photodiode (PDB435A, Thorlabs) together with the reference beam to eliminate the perturbation from the laser itself. The output from the balanced photodiode and the driven signal of the acousto-optic modulator are fed to a digital signal processing lock-in amplifier (OE1022, SYSU Scientific Instruments) for demodulation. The demodulated signal then enters into a proportional-integral servo controller (LB1005, New Focus) as an error signal for frequency stabilization, which can be viewed on an oscilloscope, and the output of the servo controller is fed back to control the piezoelectric transducer inside the laser resonator.

In the experiment, the laser wavelength of 739 nm used for locking is close to certain rovibrational transition lines of iodine molecule, which need to be heated to a high  temperature to stimulate strong enough transition activity. However, the high temperature will increase the concentration of iodine molecules, thus increasing the Doppler broadening and collision broadening effects. So we customized a 25 cm long iodine cell with a 10 cm condensation finger, which needed to be vacuumed to reduce interference from other gas molecules. During the experiment, the temperature of the iodine gas cell should be heated to 500 $^\circ$C, while the temperature of condensation finger should be kept around 36.4 $^\circ$C, at which the saturated vapor pressure of iodine is about 99.6 Pa.

The iodine vapor pressure can be calculated by the fitted formula from the reference data. \cite{haynes2014crc}:
\begin{equation}
\begin{aligned}
\log_{10}P=&-18.512+0.097360*(T+273.15)\\
&-1.0048\times 10^{-4}\times (T+273.15)^{2},
\end{aligned}
\end{equation}
where $P$ is the iodine vapor pressure in Pascals, and $T$ is the cold-finger temperature in Celcus.

\begin{figure}[H]
\begin{center}
\includegraphics[width=1.0\linewidth]{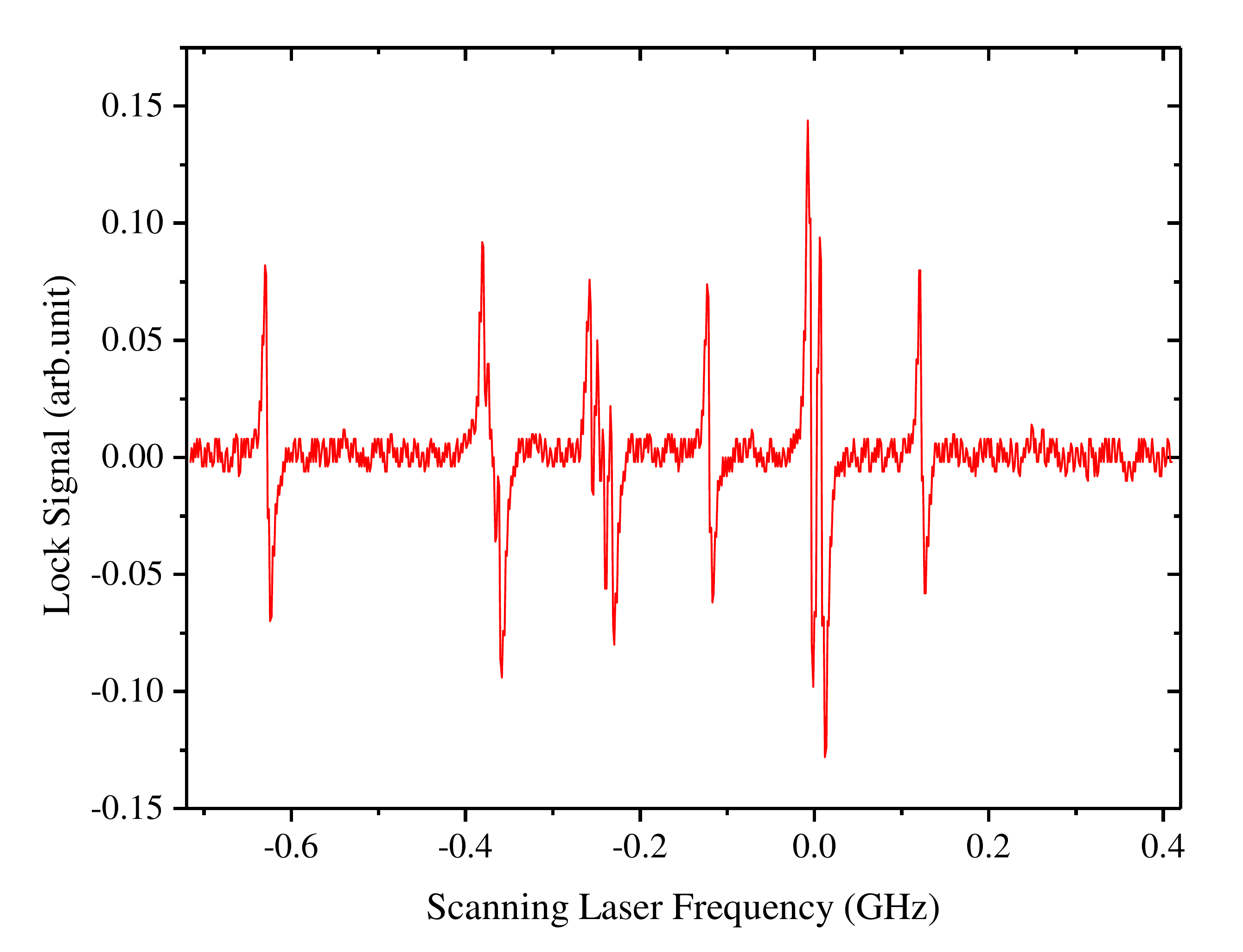}
\caption{Measured $I_2$ hyperfine transitions of the B(1)-X(11) P(70) line using modulation transfer spectroscopy. }
\label{MTS}
\end{center}
\end{figure}

\subsection{MTS spectroscopy}

We have obtained modulation transfer spectrum using counter-propagating pump and probe beam, which is shown in Fig. \ref{MTS}. In the experiment, the RF power supply to the fiber-EOM is turned off, the transmitted 8 mW is used. After the double pass of AOM, the power of the pump beam is about 2.9 mW. And the power of the probe beam is 320 $\mu$W. The frequency of the laser is scanned by the tuning the laser resonator about 1 GHz range. It has six absorption peaks near 739.0289 nm, the stronger peak of the doublet feature at the fifth peak is used to stabilize the laser frequency.By taking iodine molecule as a vibrating rotator and using PGOPHER program to simulate its spectrum  \cite{western2017pgopher}, we have assigned, for the first time, these obtained transitions to the hyperfine components of the B(1)-X(11) P(70), as Figure 3 shows.

\begin{figure}[H]
\begin{center}
\includegraphics[width=1.0\linewidth]{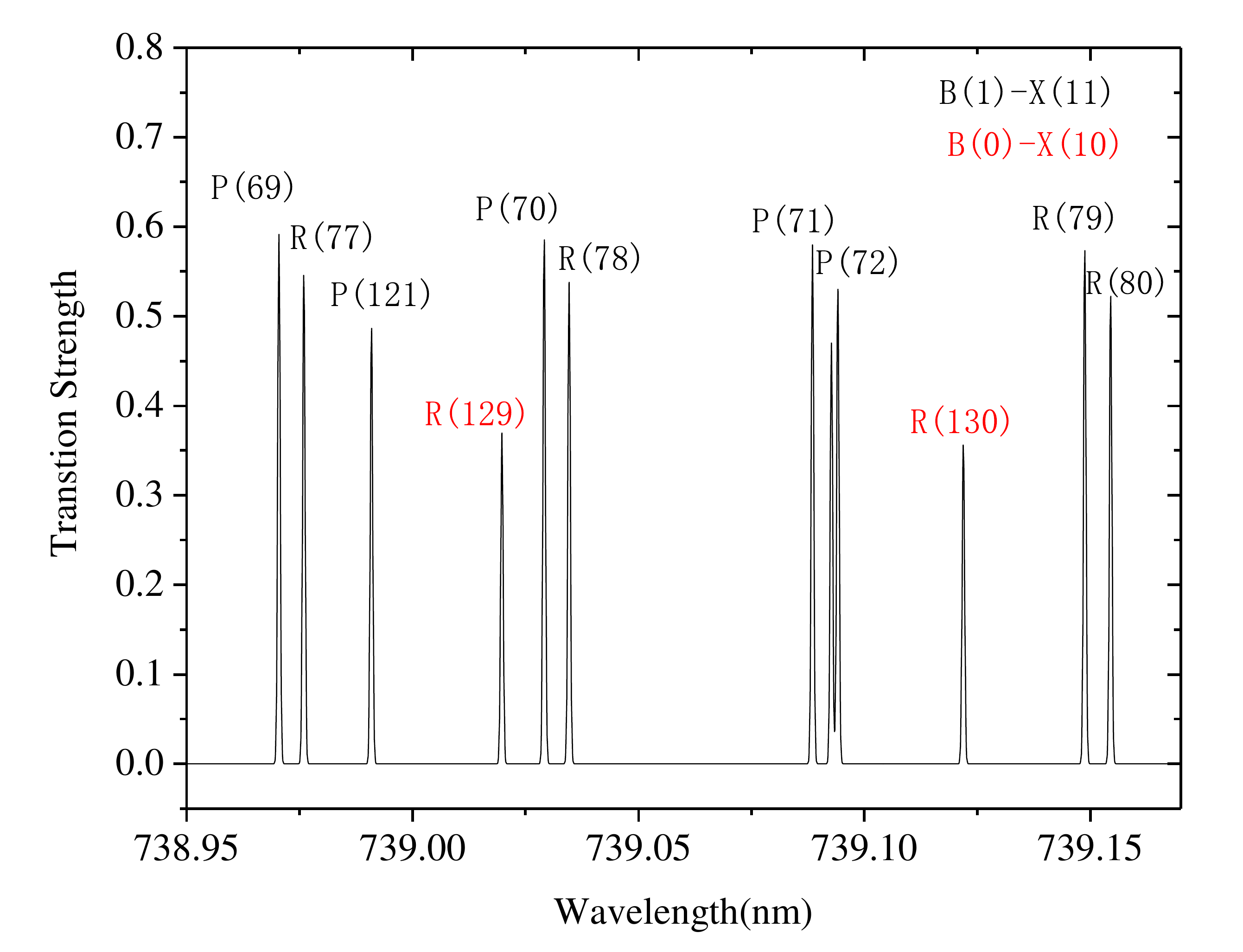}
\caption{The simulated result of the iodine absorption spectrum by PGOPHER program around 739 nm.  }
\label{MTS}
\end{center}
\end{figure}

In order to investigate the chosen locking transition, the power broadening and pressure broadening effects are studied. Figure 4 shows the linewidth (full width at half maximum, FWHM) of the locking transition increases with the pump power from 1 mW to 30 mW at four different cold finger temperatures 41.6$^{\circ}$C, 36.4$^{\circ}$C and 31.7$^{\circ}$C and 26.5$^{\circ}$C. In these experiments, the laser beam from the Ti:sapphire laser does not couple into the fiber-EOM to avoid the power limit and reach the saturation level. The probe power is kept at 400 $\mu$W. The power broadening follows the relation of  \cite{demtroder2013laser}

\begin{figure}
\begin{center}
\includegraphics[width=1.0\linewidth]{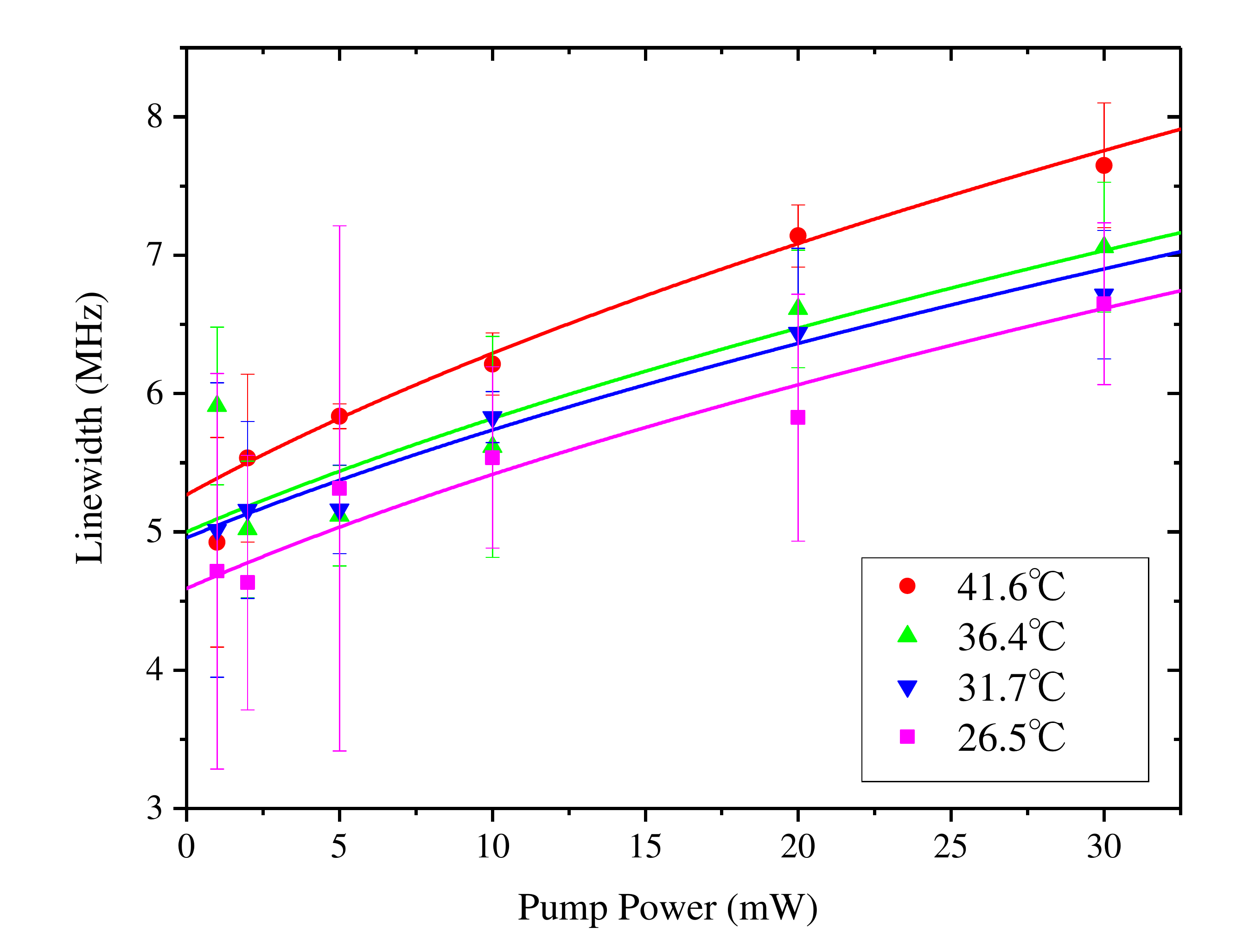}
\caption{Power broadening at the locking hyperfine component of the B(1)-X(11) P(70) transition at different cold finger temperatures. Circles (red), up triangles (green), down triangles (blue) and cubes (magenta) correspond to cold finger temperatures of the iodine cell fixed at 41.6$^{\circ}$C, 36.4$^{\circ}$C, 31.7$^{\circ}$C, and 26.5$^{\circ}$C, respectively. The solid lines represent the fitting to Eq.\ref{pump}. }
\label{pump_power}
\end{center}
\end{figure}

\begin{equation} \label{pump}
\gamma_s = \frac{\gamma}{2} (1+\sqrt{1+\frac{P_{pump}}{P_s}})
\end{equation}

where $\gamma$ is the linewidth associated to the iodine vapor pressure (or cold finger temperature), transit-time, and laser linewidth etc. $P_{pump}$ is the power of the pump beam, and $P_s$ is the saturation power. A nonlinear least-squares fit to the Eq. \ref{pump} at several cold finger temperatures is applied for the locking hyperfine component of the B(1)-X(11) P(70) transition. At each temperature, power broadening is removed by extrapolating the linewidth to zero power, thus we obtain $\gamma$ for all four temperatures and $P_{s} = 12.4\pm1.3$ mW. Then, the zero-power linewidths at different iodine pressures are fitted to a quadratic pressure dependence \cite{chen2005high}

\begin{equation} \label{pressure}
\gamma = \gamma_{0} + aP + bP^2.
\end{equation}
The dependence of $\gamma$ on the iodine vapor pressure and its fitting to Eq 3 are shown in Figure \ref{pressure}. From the pressure fit, we have $\gamma_{0} = 3.890$ MHz, $a = 19.3$ kHz/Pa, and $b = -67.35$ Hz/Pa$^2$.

\begin{figure}[H]
\begin{center}
\includegraphics[width=1.0\linewidth]{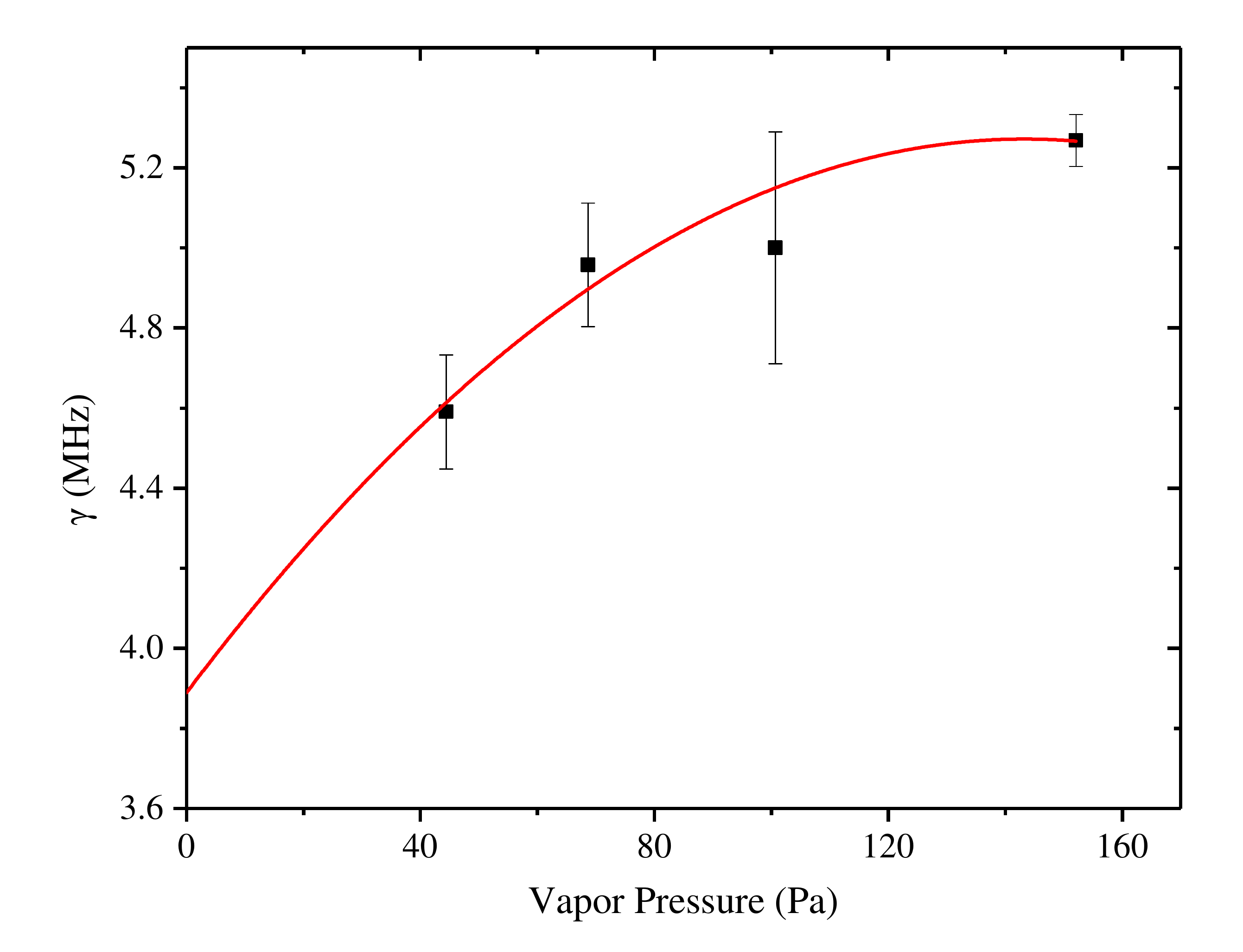}
\caption{Pressure broadening of the locking hyperfine component of the B(1)-X(11) P(70) transition. The power broadening at each pressure has already been removed by extending the pump power to the zero limit in Fig 4. }
\label{pressure}
\end{center}
\end{figure}

\subsection{Frequency stability}
We analyze the results of the MTS spectroscopy for the frequency stabilization condition. Due to the power limit of the fiber-EOM and the weak iodine rovibrational transition at 739 nm, the ratio of the lock signal to the linewidth (i.e. TSNR \cite{hopper2008optimizing})is experimentally verified to be optimal around 100 Pa, corresponding to the cold finger temperature at 36.4$^{\circ}$C.
We use the wavemeter (WS7-30, HighFinesse) to monitor the real-time stabilized frequency to evaluate the frequency drift of the locked laser. The state-of-the-art wavemeters often cover a wide operating range, and have resolution and accuracy at a level of a few MHz. We calculate the Allan deviation \cite{riley2008handbook}

\begin{equation}
\sigma_y (\tau)=\sqrt {\frac{1}{2(M-1)}\sum_{i=1}^{M-1}(y_{i+1}-y_{i})^2},
\end{equation}
where $y_i$ is the $i$th of $M$ fractional frequency values averaged over the measurement interval $\tau$. Figure 5 shows that the stability of the stabilized laser reaches a level of $3.83\times 10^{-11}$ around 13 s integration time. The inset of Figure 5 directly shows the long-term drift of the 739 nm laser in locking condition, which is less than 2.75 MHz in 7.5 hours.

\begin{figure}
\begin{center}
\includegraphics[width=1.0\linewidth]{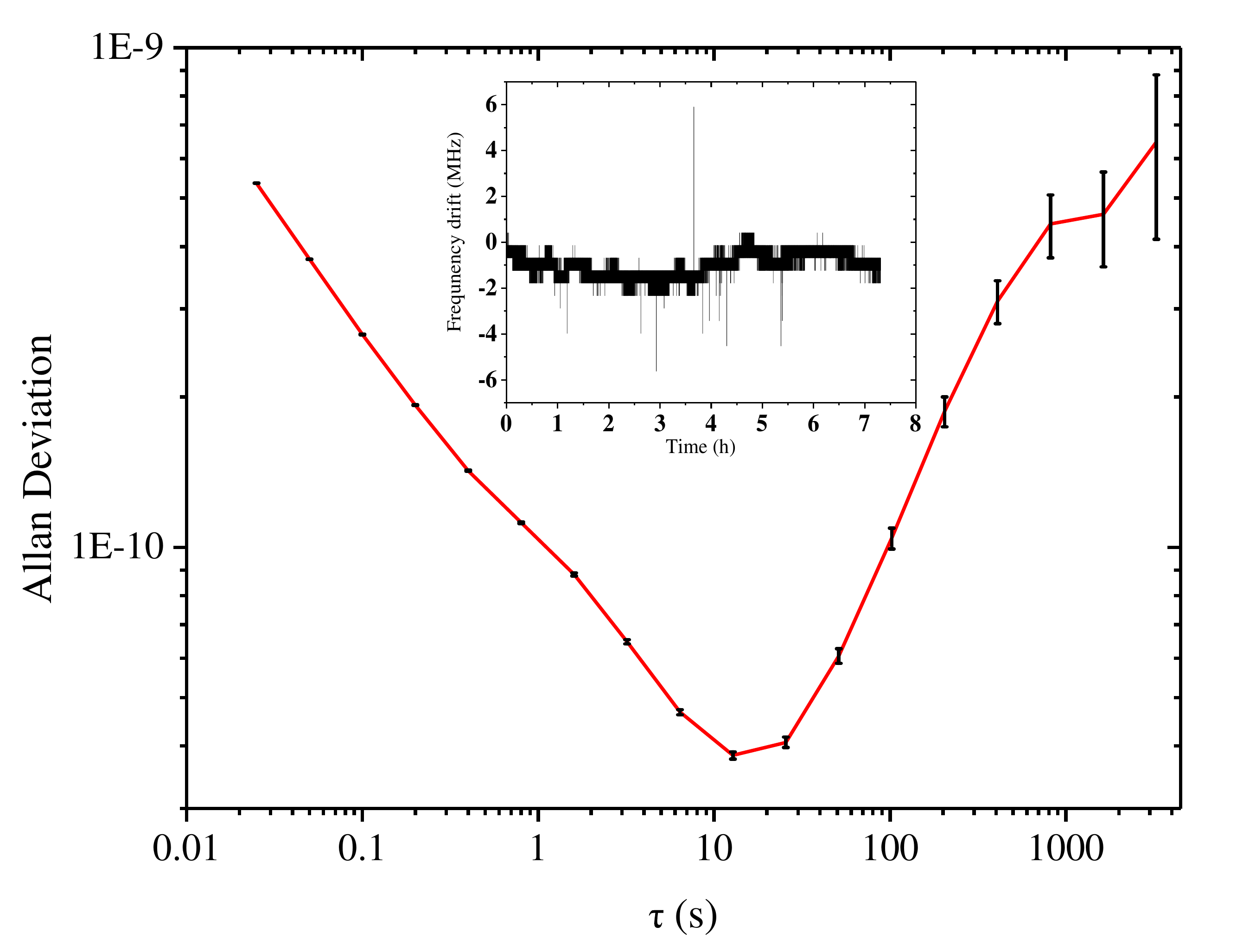}
\caption{Allan deviation $\sigma (\tau)$ of the locked 739 nm laser. The stability of the stabilized laser reaches a level of $3.83\times 10^{-11}$ around 13 s integration time. The inset directly shows the long-term drift of the 739 nm laser in the frequency locking condition which is less than 2.75 MHz in 7.5 hours. }
\label{Allan}
\end{center}
\end{figure}

\subsection{$^{171}Yb^+$ Spectroscopy}

\begin{figure}
\begin{center}
\includegraphics[width=1.0\linewidth]{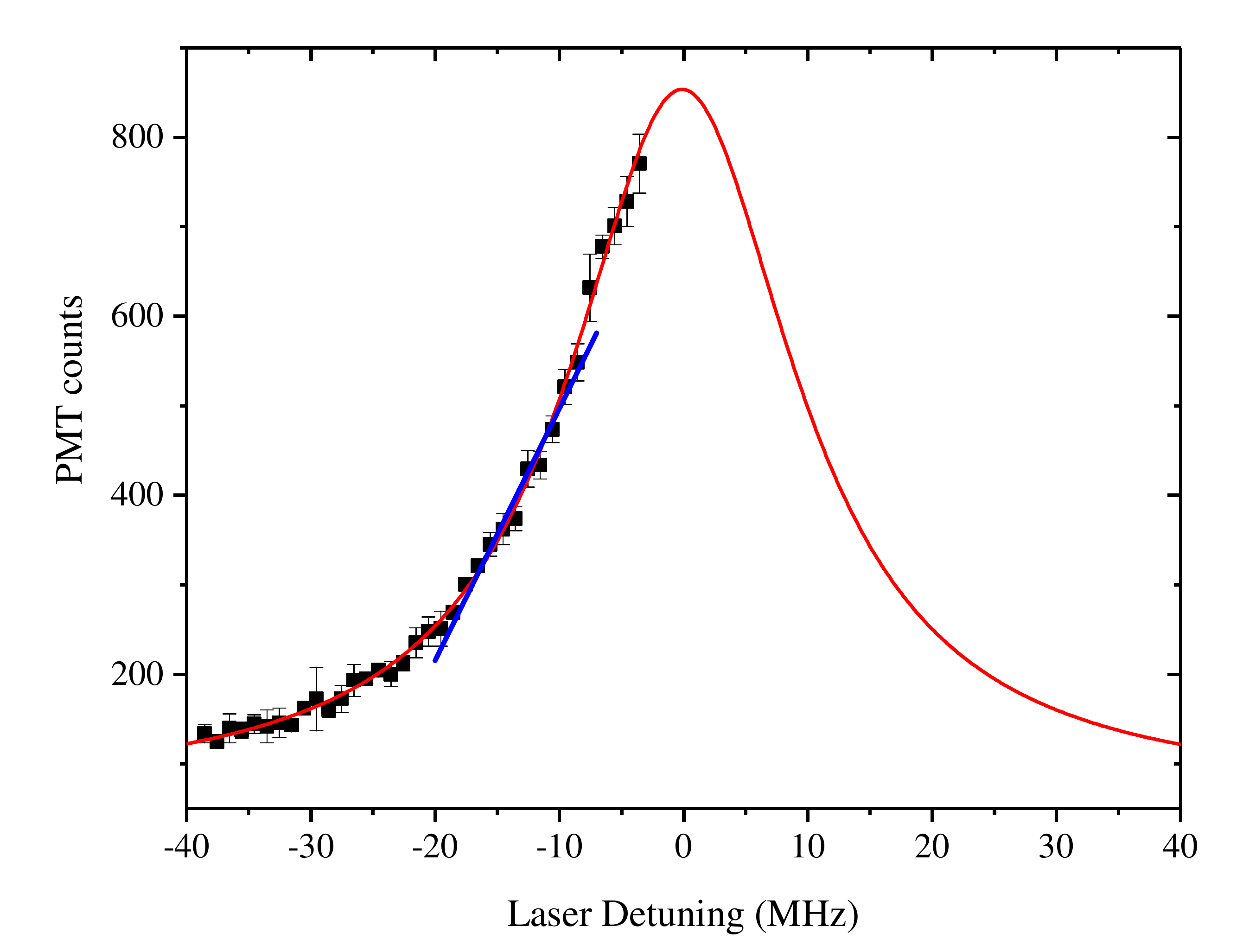}
\caption{Fluorescence excitation spectrum of a single $^{171}Yb^{+}$ ion by scanning the frequency of the 369 nm laser. Each data point corresponds to 1 ms PMT integration time. The 369 nm laser power is 750 nW. The red curve represents the fit with Lorentz profile with a FWHM of 22.2 MHz. The blue line is a linear fit to the points near $f^{\prime}= -13.6 $ MHz detuning, showing a slope of  $k=2.8138\times 10^4$ counts/s/MHz. }
\label{spectrum}
\end{center}
\end{figure}

With $\Delta f = 13.0400$ GHz frequency shift set by the fiber EOM, the 739 nm laser is locked to $f=405657.4728$ GHz of a hyperfine component of iodine. The second harmonic frequency of the 739 nm laser corresponds to $f^{\prime}= -13.6 $ MHz detuning from $^{2}S_{\frac{1}{2}}-^{2}P_{\frac{1}{2}}$ transition. In the experiment we scan the frequency of the 369.5 nm laser by tuning the fiber EOM, and obtain a spectrum as shown in Fig. \ref{spectrum}. The 369.5 nm laser power is kept at 750 nW. The spectrum is fitted to a Lorentz profile, which gives a FWHM of 22.2 MHz, indicating saturation power at 1.13 $\mu$W according to Eq. \ref{pump}.

We can also characterize the absolute frequency instability of 739 nm laser by monitoring the fluctuations in the ion fluorescence rate during the steady-state cooling process, since the fluorescence rate is dependent on the detuning of the laser frequency from the resonance of the cooling transition \cite{ghadimi2020multichannel}. We use the linear fit in Fig. \ref{spectrum} to calibrate the linear correlation between ion fluorescence count rate and laser frequency. The linear fit has a slope of $k=2.8138\times 10^4$ counts/s/MHz, which is then converted to frequency deviation of the 369.5 nm laser which can be further halved to infer the frequency deviation of the 739 nm laser.

\section{Conclusion}

The B(1)-X(11) P(70) transition of $I_2$ molecule was recorded by high-resolution acousto-optic modulation transfer spectroscopy. By optimizing the experimental conditions, we have stabilized a 739 nm laser to one of the hyperfine component of the $^{127}I_{2}$ B(1)-X(11) P(70) transition. A frequency stability of $3.83\times 10^{-11}$ around 13 s averaging time is achieved when its frequency is stabilized to the hyperfine component. The line broadening effect due to the iodine pressure and laser power are investigated in order to optimize the TSNR of the locking transition.The observed hyperfine transition of the molecular iodine is shown to be an ideal frequency reference for the cooling and detection laser for trapped ytterbium ions.

\section{Funding}

National Natural Science Foundation of China(NSFC) (11774436, 11974434, and 12074439); Natural Science Foundation of Guangdong Province (2020A1515011159); Guangdong Province Youth Talent Program (2017GC010656); Sun Yat-sen University Core Technology Development Fund, and the Key-Area Research and Development Program of GuangDong Province (2019B030330001), Fundamental Research Funds for the Central Universities of Education of China (191gpy276).


\bibliography{sample}

\bibliographyfullrefs{sample}


\end{document}